\documentclass[aps,pre,superscriptaddress,showkeys]{revtex4}

\usepackage{graphicx}
\usepackage{epstopdf}
\usepackage{amssymb,amsmath}
\usepackage{color}

\begin{document}


\title{Controlling interactions in supported bilayers from
weak electrostatic repulsion to high osmotic pressure}

\author{Arnaud Hemmerle}
\author{ Linda Malaquin}
\affiliation{UPR 22/CNRS, Institut Charles Sadron, Universit\'e de Strasbourg, 23 rue du Loess, BP 84047 67034 Strasbourg Cedex 2, France}
\affiliation{CEA, IRAMIS, SIS2M, LIONS, UMR 3299 CEA/CNRS, CEA-Saclay b\^at. 125, F-91191 Gif-sur-Yvette Cedex, France}
\author{Thierry Charitat}
\email[Corresponding author: ]{thierry.charitat@ics-cnrs.unistra.fr}
\affiliation{UPR 22/CNRS, Institut Charles Sadron, Universit\'e de Strasbourg, 23 rue du Loess, BP 84047 67034 Strasbourg Cedex 2, France}
\author{Sigol\`ene Lecuyer}
\altaffiliation{Present address: Laboratoire Interdisciplinaire de Physique, UMR, 140 avenue de la physique, Universit\'e Joseph Fourier, and CNRS, 38402 Saint Martin d'Heres, France}
\affiliation{UPR 22/CNRS, Institut Charles Sadron, Universit\'e de Strasbourg, 23 rue du Loess, BP 84047 67034 Strasbourg Cedex 2, France}
\author{Giovanna Fragneto}
\affiliation{Institut Laue-Langevin, 6 rue Jules Horowitz, BP 156, 38042 Grenoble Cedex, France}
\author{Jean Daillant}
\altaffiliation{Present address: Synchrotron SOLEIL, L'Orme des Merisiers, Saint-Aubin, BP 48, F-91192 Gif-sur-Yvette Cedex, France}
\affiliation{CEA, IRAMIS, SIS2M, LIONS, UMR 3299 CEA/CNRS, CEA-Saclay b\^at. 125, F-91191 Gif-sur-Yvette Cedex, France}

\begin{abstract}
Understanding interactions between membranes requires measurements
on well-controlled systems close to natural conditions,
in which fluctuations play an important role. We have determined,
by grazing incidence X-ray scattering, the interaction potential
between two lipid bilayers, one adsorbed on a solid surface and the
other floating close by. We find that interactions in this highly
hydrated model system are two orders of magnitude softer than in
previously reported work on multilayer stacks. This is attributed to
the weak electrostatic repulsion due to the small fraction of ionized
lipids in supported bilayers with a lower number of defects. Our
data are consistent with the PoissonÐBoltzmann theory, in the regime
where repulsion is dominated by the entropy of counter ions.
We also have unique access to very weak entropic repulsion potentials,
which allowed us to discriminate between the various models
proposed in the literature. We further demonstrate that the interaction
potential between supported bilayers can be tuned at will by
applying osmotic pressure, providing a way to manipulate these
model membranes, thus considerably enlarging the range of biological
or physical problems that can be addressed.
\end{abstract}
\date{\today}

\maketitle

\section{Introduction} Supported lipid bilayers offer a unique configuration whereby
a single bilayer, accessible to other molecules such as, for
example, proteins, peptides, or DNA, is supported on a solid
substrate. Beyond their interest for biosensor technology, the
access they give to a flat immobilized membrane makes them
highly relevant for fundamental studies in biophysics and membrane
biology \cite{sackSci96,Castellana(surfacescirep2006)}. 
particular, they provide a unique way to
finely characterize the interactions between membranes and their
environment, which are not only crucial for membrane fusion and
trafficking, endocytosis, and exocytosis \cite{Mouritsen1998a,liporevue},
but also fascinating from the physical point of view. \\
Membranes indeed exhibit extremely complex interactions
with their environment, in which both molecular-scale enthalpic
and fluctuation-related entropic contributions are inextricably
involved. In particular, the effect of confinement has been now
discussed for 40 years without a definitive answer being found.
Helfrich first realized that, in addition to the ÒdirectÓ electrostatic,
van der Waals, and hydration forces \cite{liporevue},the long-range ÒeffectiveÓ
steric interaction generated by the thermal fluctuations of confined
flexible membranes is an essential contribution to the total
free energy of interaction \cite{helfrich73}.
Pure hard wall interaction (hard confinement) was first considered in \cite{helfrich73, seifert(prl1995)} 
but is not a realistic description of real systems, and especially not of living ones. 
Confinement by a ``soft'' potential was treated either by using self-consistent methods leading to effective 
exponentially decaying potentials \cite{Evans1986, sornette:4062, Podgornik1992}, or by estimating average values within a full statistical mechanics approach \cite{Mecke2003}.
Which functional form
should be used to describe entropic repulsion in real experimental
situations, however, remains an open question.

\begin{figure}[h]
\begin{center}
\includegraphics[width=10cm]{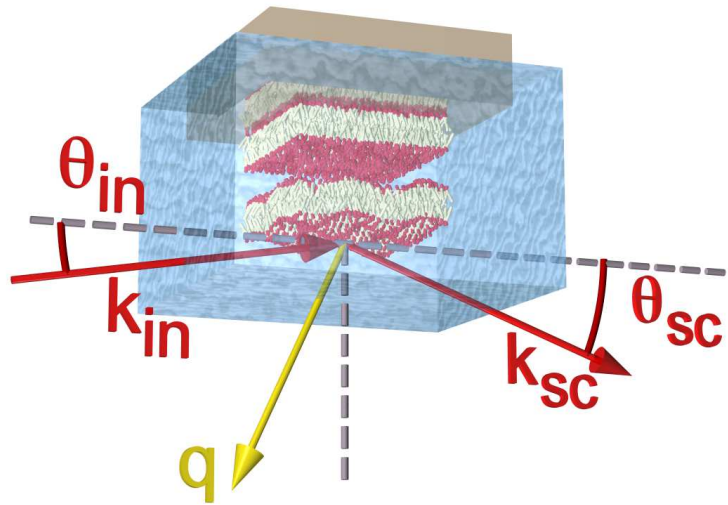} 
\caption{Schematic view of the experimental setup for specular and off-specular reflectivity. The grazing and scattered wavevectors (resp. angles of incidence)
are ${\bf k}_{in}$ and ${\bf k}_{sc}$ (resp. $\theta_{in}$ and $\theta_{sc}$). ${\bf q}$ is the wavevector transfer.} 
\end{center}
\label{figure1} 
\end{figure}

Although the surface force apparatus can be used to precisely
determine the direct part of the potential (hydration and van der Waals contributions) \cite{rand(bba1989)},the entropic repulsion can only be
studied by using scattering techniques. Combination with osmotic
pressure measurements allowed in particular the determination
of the compressibility $B=\partial^2{\cal F}/\partial d_w^2$, where ${\cal F}$ is the system free energy and $d_w$ the interlayer water thickness \cite{rand(bba1989),NagleBBA2000}. Although some agreement was found with the soft potential
of Ref.\cite{Podgornik1992}, the experimental decay lengths found in \cite{petrache(pre1998)} were greater than twice the value predicted by theory, $\sim 0.2$ nm. Moreover, inconsistencies between the temperature dependence of $B$ and the observation of an unbinding transition pointed to the role of static defects in multilayers, which would dramatically affect bilayer interactions \cite{Salditt2000}. 
In the work described here, we determine the interaction potential between supported bilayers. These bilayers can be almost defectless but contain much less material than multilayers and could not be studied using diffuse scattering until recently \cite{daillant2005}.

\section{Results and discussion}

\begin{figure}[ht]
\includegraphics[width=10cm]{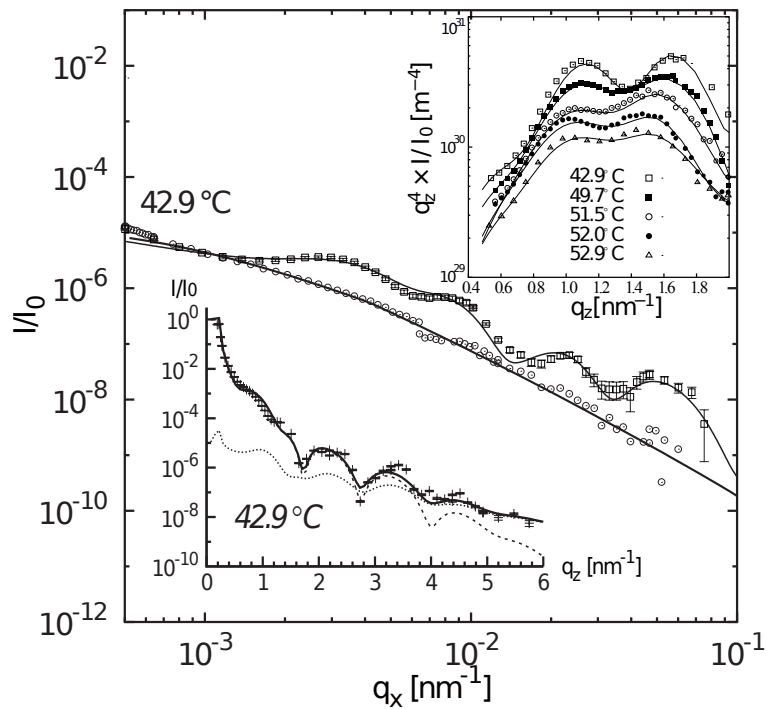} 
\caption{Off-specular reflectivity from silicon substrate ($\circ$) and an OTS bilayer at T=42.9$^\circ$C ($\square$) as a function of $q_x$.
Continuous lines represent best fits.
Top inset: off-specular reflectivity as a function of $q_z$ zoomed in the region where it is most sensitive to the potential 
at T=42.9$^\circ$C ($\square$), 49.7$^\circ$C ($\blacksquare$), 51.5$^\circ$C ($\circ$), 52.0$^\circ$C ($\bullet$), 52.9$^\circ$C ($\triangle$). 
Note the shift in minimum and decrease in contrast with increasing temperature.
Bottom inset: specular reflectivity (continuous line), ``true'' specular reflectivity (dashed line), and diffuse scattering in the specular direction $q_x=0$ (dotted line) calculated using the model of Ref. \cite{malaquin10} using the fit parameters and experimental resolution.
} 
\label{figure2} 
\end{figure}

\begin{figure}[h]
\includegraphics[width=11cm]{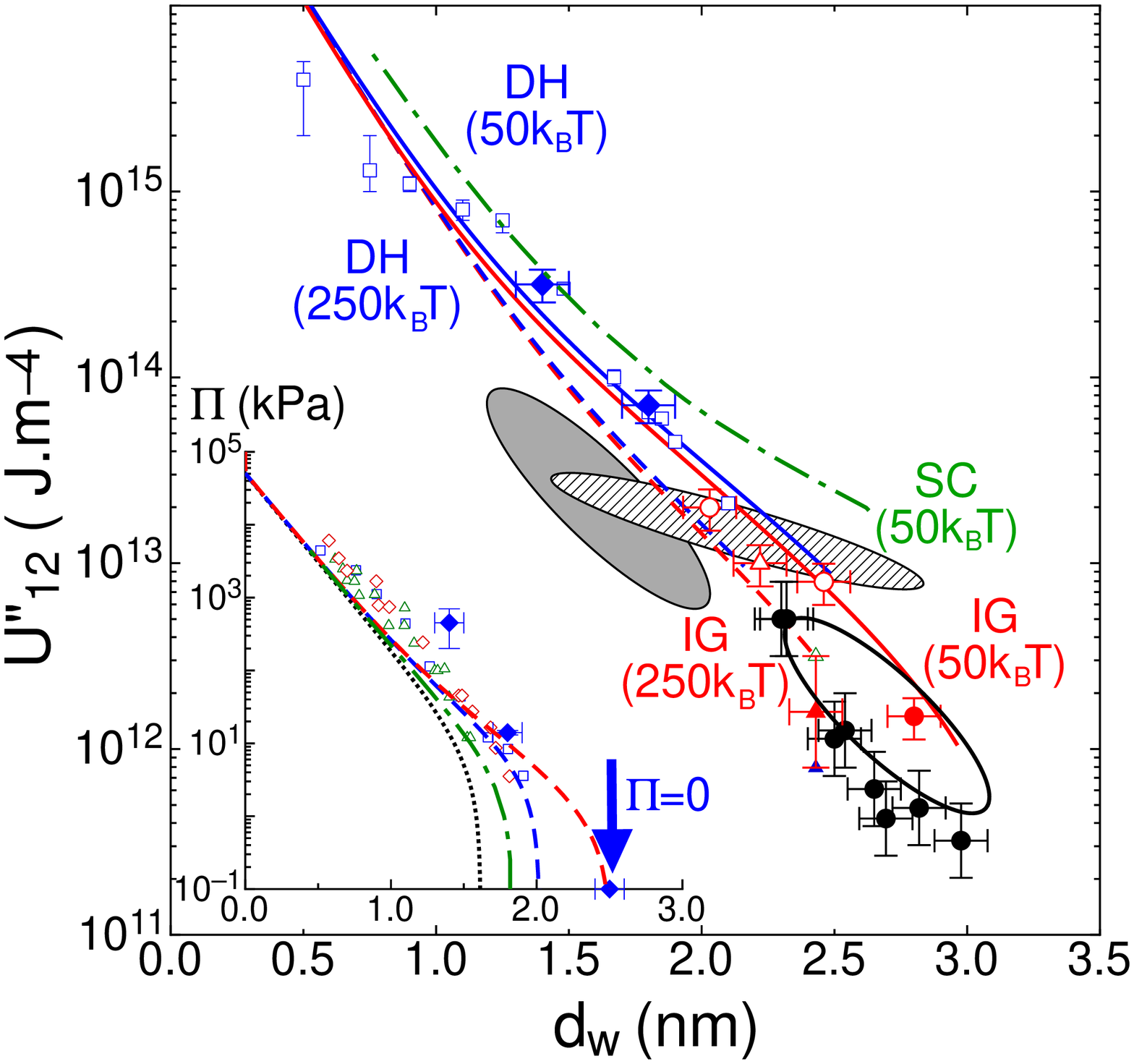}
\caption{Interaction potential second derivative $U^{\prime\prime}_{12}$ as a function of the interlayer water thickness $d_w$. ($\textcolor{blue}{\square}$): data from Petrache et al \cite{petrache(pre1998)} for Egg PC in fluid phase. All other data from this work using DSPC: fluid phase ($\bullet$); fluid phase prior to ($\textcolor{red}{\bullet}$) and after($\textcolor{red}{\circ}$) the addition of salt (T=58$^\circ$C; $\ell_D=0.5$ nm and $0.4$ nm);  gel phase prior to ($\textcolor{red}{\blacktriangle}$) and after ($\textcolor{red}{\triangle}$) the addition of salt (T=40$^\circ$C; $\ell_D=0.4$ nm); ($\textcolor{blue}{\blacklozenge}$) Gel Phase with applied osmotic pressure ($\Pi =10 \pm 4$ and $450 \pm 250$ kPa). 
Red solid (respectively dashed) lines: soft-confinement potential \cite{Podgornik1992} plus electrostatic contribution in the Ideal Gas limit (IG) for Fluid (respectively Gel) phase. 
Blue solid (respectively dashed) lines: soft-confinement potential \cite{Podgornik1992} plus electrostatic contribution in the Debye-H\"uckel limit (DH) ($\ell_D=0.3$ nm) for Fluid (respectively Gel) phase. Green dashed-dotted line: self-consistent model \cite{Mecke2003} in fluid phase ($\kappa=50 k_B T$).
The ellipses show the region of equilibrium states without applied pressure for Helfrich, hydration and van der Waals forces (grey area); soft-confinement, hydration and van der Waals forces (dashed area); soft-confinement, hydration and van der Waals forces plus electrostatic interaction in the ideal gas limit (empty ellipse). 
The different ellipses were obtained by varying $P_h$, $z_h$, $d_{\rm head}$ and $\kappa$ within the limits indicated in text. 
Inset: osmotic pressure $\Pi$ (interaction potential first derivative $U^{\prime}_{12}$) as a function of the interlayer water thickness $d_w$. 
($\textcolor{blue}{\square}$): Open symbols, data from Petrache et al \cite{petrache(pre1998)} for Egg PC ($\textcolor{blue}{\square}$), DMPC ($\textcolor{red}{\lozenge}$) and DPPC ($\textcolor{green}{\triangle}$) in fluid phase. ($\textcolor{blue}{\blacklozenge}$) data from this work using DSPC in gel phase with applied osmotic pressure. 
Red dashed line: soft-confinement potential \cite{Podgornik1992} with electrostatic contribution in the Ideal Gas limit for the Gel phase. 
Blue dashed line: soft-confinement potential \cite{Podgornik1992} plus electrostatic contribution 
in the Debye-H\"uckel limit (DH) ($\ell_D$=0.3 nm) for the Gel phase. 
Green dashed-dotted line: self-consistent model \cite{Mecke2003} in gel phase ($\kappa=250 k_B T$). 
Black dotted line: microscopic potential plus electrostatic contribution without any entropic contribution. 
The same set of parameters has been used in main figure and in the inset.}
\label{figure3} 
\end{figure}

Two kinds of supported bilayers were investigated in this study, both consisting of two bilayers (see Fig. \ref{figure1}). The first type, called ``double bilayers" in the following, consists of two bilayers of DSPC, whereas in the other, called ``OTS-bilayer", the first monolayer close to the substrate is replaced by an octadecyl-trichlorosilane grafted layer (see Materials and Methods for details). In both cases, the second bilayer is free to fluctuate in the potential of the first bilayer and of the substrate. 
A combined fit of specular and off-specular data is performed in order to increase sensitivity, using a model taking into account the static and thermal roughness from both bilayers (see Ref. \cite{malaquin10} and Materials and Methods).
From the fits presented in Fig. \ref{figure2}, we obtain structural parameters, in particular the bilayer-bilayer distance and the interlayer water thickness, but also the interaction potential second derivative, and the bilayer tensions and bending rigidities (Fig. \ref{figure3}). Electron density profiles for OTS or double bilayers can be found in Ref. \cite{malaquin10} (an example is given in Supporting Information).

\begin{figure}[h]
 \includegraphics[width=10cm]{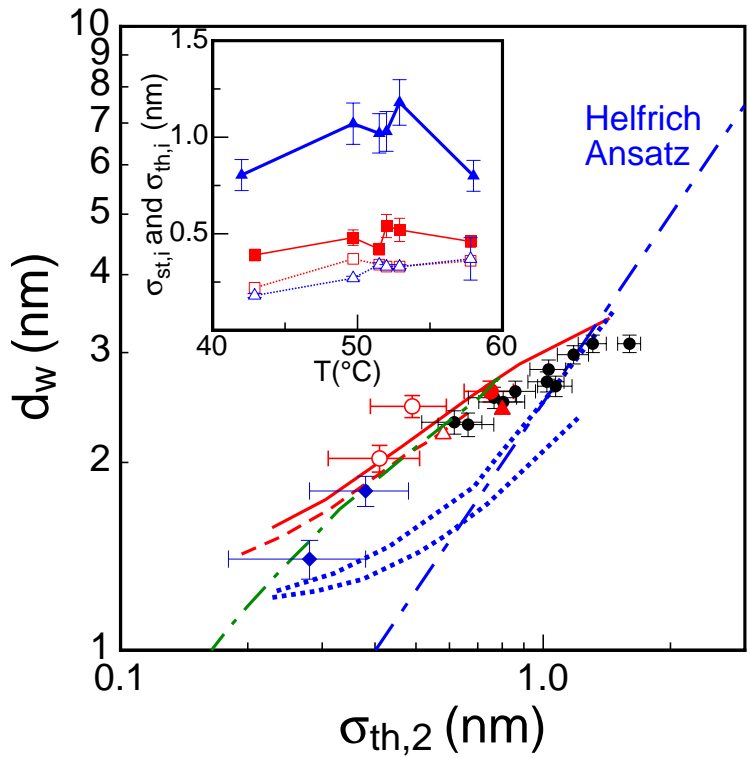}
\caption{ $d_w$ as a function of $\sigma_{th,2}$. Symbols are the same as for figure \ref{figure3}. Dashed-dotted line (blue online): ``Helfrich Ansatz'' $d_w^2 = 1/6 \sigma_{th,2}^2$. Dotted lines (blue online): Helfrich confinement, with a 0.3 nm shift for the lower curve as explained in text. Dashed line (red online): Soft potential (Ref. \cite{Podgornik1992}) without electrostatic contribution. Solid line (red online): Soft potential (Ref. \cite{Podgornik1992}) with electrostatic contribution. Green dashed-dotted line: self-consistent model \cite{Mecke2003} with electrostatic contribution. Inset: static roughness $\sigma_{st,i}$ (open symbols) and thermal roughness $\sigma_{th,i}$ (closed symbols) of the first strongly adsorbed bilayer (i=1, square, red online) and of the floating bilayer (i=2, triangle, blue online).}
\label{figure4} 
\end{figure}

In all cases, the best fit values for the floating membrane tension is $0.3 \pm 0.2$ mN/m.  The bending rigidity decreases from $(250 \pm 50) k_BT$ in the gel phase ($T=42.9^\circ$C) to $(50 \pm 20) k_BT$ in the fluid phase ($T=52.9^\circ$C) in agreement with previously reported values \cite{daillant2005}. 
The static, substrate-induced roughness of both membranes is always less than $0.3$ nm and remains constant, as well as the thermal roughness $\sigma_{th,1}$ of the first, adsorbed membrane, which is on the order of $0.4$ to $0.5$ nm 
(Fig. \ref{figure4} inset). The thermal roughness $\sigma_{th,2}$ of the second bilayer is larger than that of the first bilayer, justifying the denomination ``floating bilayer", in good agreement with previous experiments \cite{malaquin10,sinha(PRE2011)}.\\
A zoom of the off-specular reflectivity in the region where it is most sensitive to the interaction potential is shown in Fig. \ref{figure2} for different temperatures.
It is important to note here that the second derivative of the interbilayer interaction potential is directly linked to the depth of the minimum in the diffuse scattering curve around $q_z \approx 1.0-1.5$ nm$^{-1}$, without much coupling to the other parameters. 
Similarly, the interlayer water thickness is strongly correlated to the $q_z$ position of that minimum. Hence, it can be seen directly in the inset of Fig. \ref{figure2} that the interaction potential becomes weaker (the minimum is less pronounced) when the interlayer water thickness increases (left shift of the minimum) at higher temperatures.

The second derivative of the interbilayer potential $U^{\prime\prime}_{12}$ obtained by fitting the experimental data  
is represented as a function of the interlayer water thickness $d_w$ in Fig. \ref{figure3}, where
our data are compared with values obtained by Petrache et al. for Egg PC multilayers \cite{petrache(pre1998)}. 
Remarkably, our samples are more hydrated than multilayers ($d_w$ is 0.1 to 0.5 nm larger) and interact via 
a softer interaction potential ($U^{\prime\prime}_{12}$ is smaller).  
Note that in this analysis, the interaction potential is not only strongly constrained via its 
second derivative $U^{\prime\prime}_{12}$, but also via $d_w$ which fixes the position of its minimum. 
$d_w$ is in turn reported as a function of the second bilayer thermal roughness $\sigma_{th,2}$ 
in Fig. \ref{figure4} which shows a strong correlation between the two parameters, demonstrating that there is a large entropic contribution to the repulsion as expected.\\
The accuracy of our data, in particular for large separations, allows for a precise assessment of the repulsive part of the potential. 
In particular, it allows for a test of the different functional forms which have been used in the literature to model the entropic part of the potential.
With this aim, we first accurately calculated the attractive part of the potential using the Lifshitz theory \cite{Shubin1993}, carefully modelling the silicon-silicon oxide-water-lipid bilayer-water-lipid bilayer stack (for more details see Supporting Information). 
For $d_w < 3$ nm, a good approximation to the van der Waals interaction energy is $U_{\rm vdW} = -H/12 \pi (d_w+2d_{\rm head})^2$, with
$H=5.3\times 10^{-21}$ J and $d_{\rm head}$ the headgroup layer thickness ($0.4-0.8$ nm) in good agreement with previous work \cite{marra(biochemistry1985),Parsegian1993}.\\
We further modeled hydration forces using a classical exponential decay, $U_{\rm hyd} = P_h z_h \exp{(-z/z_h)}$, 
with $P_h = 1-5 \times 10^7$ Pa the hydration pressure and $z_h = 0.16-0.2$ nm the hydration length \cite{petrache(pre1998)}.\\
The renormalization of the microscopic interaction potential by the thermal fluctuations is a complex problem of modern statistical physics. 
According to Helfrich \cite{helfrich78}, the membrane free energy is the sum of the microscopic potential and of the entropy cost of confining the flexible membrane, which results in an effective potential controlling bilayer position. 
The effective potential, average bilayer position, and fluctuation amplitude are thus coupled quantities which must be self-consistently determined.\\
As mentioned in the introduction, Podgornik and Parsegian have extended Helfrich's approach \cite{helfrich78} to take into account hydration repulsion and van der Waals attraction in the so-called  ``soft'' potential \cite{Podgornik1992}, and 
the self-consistent approaches of Refs. \cite{Mecke2003, manghi(langmuir2010)} allow the calculation of average position, rms roughness and mean effective potential curvature.
All these theories are based on a quadratic approximation of the interaction potential, either symmetric \cite{helfrich78} or not \cite{Podgornik1992, Mecke2003, manghi(langmuir2010)}. 
In any case, a non-symmetric case, like ours, can always be mapped to a symmetric case by identifying the strength of the quadratic potential with the second derivative of the asymmetric potential. 
As an example, the Helfrich potential has been shown to correctly describe a supported bilayer interacting with a single hard wall in the limit of small fluctuations (see \cite{Mecke2003} and Supporting Information).

We first considered hard wall repulsion, with $U_{\rm Hel} = c_H /\kappa (k_BT/z)^2$ \cite{helfrich78} per unit area  
and $c_H$ ranging from 0.08 to 0.2 \cite{Gompper1989,Mecke2003}. The corresponding $d_w$ equilibrium values at zero applied pressure are given by the dark grey area on Fig. \ref{figure3} and show poor agreement with the data. 
As expected, the hard-wall potential also fails to describe the $d_w$ vs $\sigma_{th,2}$ curves (Fig. \ref{figure4}).
Accordingly, the classical ``Helfrich Ansatz'' $d_w \propto \sigma_{th}$ does not apply (Fig. \ref{figure4}). 
We also note that simply shifting the zero of the potential to account for softness does not help and that the potential of 
Ref. \cite{seifert(prl1995)} taking membrane tension into account cannot be distinguished from the hard-wall potential for realistic tension values.
The ``soft'' potential of Podgornik and Parsegian \cite{Podgornik1992} $U_{\rm soft} = \pi k_B T /16 \sqrt{P_h/\kappa/z_h} \exp(-z/n z_{h})$ 
leads to a slightly better agreement but still predicts $U^{\prime\prime}_{12}$ values one to two orders of magnitude larger than those observed experimentally (Fig. \ref{figure3}) and also underestimates the values of $d_w$ and $\sigma_{th,2}$ data (Fig. \ref{figure4}). Finally, the self consistent theory of Ref. \cite{Mecke2003}, which in principle allows one to calculate the mean $d_w$, $U^{\prime\prime}_{12}$ or $\sigma_{th,2}$ more satisfactorily than effective potential theories, is indeed in good agreement with the $d_w$ vs $\sigma_{th,2}$ data,  but it strongly overestimates $d_w$, $\sigma_{th}$ and $U^{\prime\prime}_{12}$, probably because of a bad sampling of the most confined microstates.

As our bilayers weakly interact with the substrate and can be very close to detachment, which was sometimes observed for temperatures $\gtrsim 60 ^{\circ}$C, a very small but long-range repulsive contribution to the potential would in fact be enough to shift  the equilibrium position to higher $d_w$ and lower $U^{\prime\prime}_{12}$. 
Such electrostatic interactions are investigated in Ref.\cite{parsegian(biochem1978)}, but always ignored in scattering studies on zwitterionic lipids \cite{petrache(pre1998)}.
Phosphatidylcholines present indeed $pKa$ values of 2.7 and 11 \cite{sanden10} and bear a positive charge density  $\sigma \sim 0.001$ e$^-$/nm$^2$ at the experimental $pH=5.5$.
This small amount of charges, necessarily present due to the amphoteric character of the phosphatidylcholine group, leads to a weak electrostatic repulsion which was recently shown sufficient to prevent vesicles from adhering \cite{pincet(EPJB1999)}.\\

In our experiments, surface charge densities are small and the mean-field Poisson-Boltzmann (PB) theory is expected 
 to appropriately describe the system. For DSPC supported bilayers in ultra-pure water, the Gouy-Chapman length related to the surface charge density 
$\sigma$ is $\ell_G \sim 10^2-10^3$ nm, the Debye-H\"uckel length which describes screening is $\ell_D \sim 200-500$ nm and $d_w \sim 2-3$ nm. 
In this regime, the Ideal Gas (IG) limit of the mean-field Poisson-Boltzmann theory, where the interaction potential reads $U_{el} = 2 k_BT \sigma/e \ln(d_w)$, should apply rather than the Debye-H\"uckel limit (DH) \cite{andelman(review1995)}. 
Adding this contribution to the soft potential of Ref. \cite{Podgornik1992} gives a perfect description of our data for both representations $U^{\prime\prime}_{12}$ vs $d_w$ (Fig. \ref{figure3}) and $d_w$ vs $\sigma_{th,2}$  (Fig. \ref{figure4}). 
Adding the electrostatic contribution extends the curves previously obtained to larger $\sigma_{th,2}$ and $d_w$ without significantly modifying their low $d_w$ part. Therefore, it cannot significantly improve the agreement for the other forms of the potential (Fig. \ref{figure3}). 
Accordingly, the charge density is very robust and does not depend on the other parameters. 
The validity of the soft effective potential in the presence of electrostatic interactions could nevertheless be questioned as the soft potential was explicitely constructed for hydration and van der Waals forces only. 
Though the influence of electrostatic interaction on entropically induced repulsive forces has not been investigated in detail, these are very long range compared to van der Waals and hydration interactions, and should be only marginally renormalized. 
Moreover, it was shown in Ref. \cite{andelman(review1995)}, that in the absence of van der Waals and hydration forces, the renormalization of electrostatic interactions is negligible in the limit where the in-plane electrostatic correlation length $\xi$ (50-100 nm here) is larger than $d_w$.\\
In order to further check the effect of electrostatic interactions, NaCl was added to a double bilayer.
This is expected to increase screening and is easier to analyse than changing the $pH$ to change the lipids degree of ionization. 
We prepared two solutions with $\ell_D=0.45$ nm and $\ell_D=0.3$ nm (see Materials and Methods). The results are reported in Fig. \ref{figure3} and \ref{figure4}. 
A strong decrease in the interlayer water thickness $d_w$ and in the thermal roughness $\sigma_{th,2}$ is observed, as well as a large increase in the interbilayer potential $U^{\prime\prime}_{12}$, in good agreement with the strong screening of the electrostatic potential.\\

A further proof of the good control we have over the interactions between supported bilayers is provided in the inset of Fig. \ref{figure3} where the effect of osmotic pressure is shown.  Osmotic pressure was applied using PVP (see Materials and Methods for details), and here again, the agreement with the theoretical model and with previous experiments on multilamelar systems  is perfect. This is a first demonstration that the interbilayer potential of supported bilayers can be tuned using osmotic pressure, allowing us to extend the measurements towards smaller interlayer water thicknesses and to bridge the gap with multilayer studies.

\section{Conclusion} In this article, X-ray off-specular scattering measurements of the
interaction potential between two bilayers adsorbed on a solid
substrate are shown to lead to results presenting unprecedented
sensitivity, illustrated by the necessity of taking into account the
very weak electrostatic repulsion between almost neutral bilayers
and the possibility of discriminating between different entropic
and electrostatic potentials. These results show that supported
bilayers are significantly more hydrated and therefore exhibit
more intrinsic properties than the usually studied multilayers,
possibly owing to defects in the latter. This opens up a wide range
of possibilities for understanding unbinding or investigating the
effect of various biological molecules on interaction and adhesion
between membranes.



\section{Materials and Methods}

The supported bilayers were prepared by depositing two bilayers on ultra-flat silicon substrates (SESO, France), 
where the first, more strongly adsorbed, bilayer
is either a bilayer of $\scriptstyle{L{-}\alpha}$ 1,2-distearoyl-sn-glycero-3-phosphocholine
(DPSC, Avanti Polar Lipids, Lancaster, Alabama) made by a combination of classical Langmuir-Blodgett (LB) and Langmuir-Schaefer (LS) depositions
(vertical sample) \cite{charitat1999} or a mixed octadecytrichlorosilane (OTS) - lipid bilayer (OTS bilayers), 
where the OTS layer is chemically grafted on the substrate \cite{hughes2002}.
A second, ``floating" bilayer is then prepared by a LB deposition, followed by a LS deposition.
The first bilayer serves both as a spacer to reduce the interaction between the floating bilayer and the 
substrate and keep it free to fluctuate, and to investigate bilayer-bilayer interactions.
The sample is then inserted into a PTFE sample cell with 50$\scriptstyle{\mu}$m thick 
windows. The sample cell is tightly closed and transferred to an alumina box, thermalized
by a water circulation, first at 25$\scriptstyle{^\circ}$C, then heated by steps, with a feedback on the
temperature measured inside the sample cell by a PT100 resistance.\\

Experiments were performed using ultra pure water (18.2 M$\scriptstyle{\Omega}$.cm), obtained from a Millipore purification system. Dissolution of CO$\scriptstyle{_2}$ in water leads to Debye-Length values around $\scriptstyle{\sim}$200 nm, ten times smaller that the 960 nm expected for such samples \cite{Haughey1998217}. By adding Sodium Chloride $\scriptstyle{c=}$0.5 and 1 M, we obtained solutions with Debye length equal to 0.45 and 0.3 nm respectively. Osmotic pressure was applied using polyvinylpyrrolidone (PVP) of average molecular weight 40 000 (Sigma Aldrich, St. Louis, MO) mixed with milli-Q water. Solutions of 4 and 30 \% of PVP (w/w) were prepared and homogenized with magnetic stirrings overnight. The bulk water was then carefully replaced by the PVP/water solutions with syringes, taking care not to expose the bilayers to air. The values of the osmotic pressures are deduced from the PVP concentrations as calculated by Vink \cite{vink(eurpolymj1971)} and reported in \cite{mcintosh(biochem1986)}, leading to $\scriptstyle{\Pi =}$ 14$\scriptstyle{\pm}$2 kPa and 450$\scriptstyle{\pm}$250 kPa.

Specular and off-specular reflectivities were recorded using the procedure of
Ref. \cite{daillant2005}.  The experiments reported here used a 27 keV x-ray beam (wavelength
$\scriptstyle{\lambda=}$0.0459 nm) at the CRG-IF beamline of the European Synchrotron Radiation Facility
(ESRF). The scattering geometry is described in Fig. \ref{figure1}.  
The monochromatic incident
beam was first extracted from the polychromatic beam using a two-crystal Si(111) monochromator.
Higher harmonics were eliminated using a W coated glass mirror, also used for focusing.  In all
experiments, the incident beam was 500 $\scriptstyle{\mu}$m $\scriptstyle{\times}$ 18 $\scriptstyle{\mu}$m (H $\scriptstyle{\times}$ V).  The reflected
intensity was  defined using a 20 mm$\scriptstyle{\times}$ 200 $\scriptstyle{\mu}$m (H $\scriptstyle{\times}$ V) slit at 210 mm from the sample
and a 20 mm $\scriptstyle{\times}$ 200 $\scriptstyle{\mu}$m (H $\scriptstyle{\times}$ V) slit at 815 mm from the sample and recorded using a NaI(Tl)
scintillator.  Specular reflectivity was obtained by rocking the sample for each angle of incidence
($\scriptstyle{q_x}$ scans for approximately constant $\scriptstyle{q_z}$) in order to subtract the background.
Off-specular reflectivity was measured at a constant grazing angle of incidence of 0.7 mrad
below the critical angle of total external reflection at the Si-water interface (0.83 mrad), leading to variation of both $\scriptstyle{q_x}$ and $\scriptstyle{q_z}$
(see Fig. \ref{figure1}).

Optimization of slit widths allowed us to extend the in-plane wavevector transfer range $\scriptstyle{q_\parallel}$ by one order of magnitude from $\scriptstyle{2 \times 10^{6}}$ m$\scriptstyle{^{-1}}$ to  $\scriptstyle{2 \times 10^{5}}$ m$\scriptstyle{^{-1}}$compared to the experiments of Ref. \cite{daillant2005}. Our experiment is thus sensitive to the off-specular diffusion by both bilayers and not only the more strongly fluctuating one,  allowing for a precise determination of interaction potentials.\\
The differential scattering cross-section (power scattered per unit solid angle per unit incident flux) 
can be written \cite{malaquin10}: 

\begin{eqnarray}
\frac{d\sigma}{d\Omega} &\approx& r_e^2 \left\vert t(\theta_{\rm in}) \right\vert^ 2 
\left\vert t(\theta_{\rm sc}) \right\vert^ 2
\left\langle \left| \int
d{\bf r} e^{i \bf{q_\parallel}.\bf{r_\parallel}} \left[  \frac{\rho_{Si} - \rho_w}{iq_z} e^{iq_z
z_s (\bf{r_\parallel})}\right.\right.\right. + \left.\left.\left. \delta \tilde{\rho}_1(q_z) e^{iq_z z_1 (\bf{r_\parallel})) }+ \delta \tilde{\rho}_2(q_z) e^{iq_z z_2 (\bf{r_\parallel})) } \right] \right|^2
\right\rangle, 
\label{equagene} 
\end{eqnarray} 

where $\scriptstyle{r_e = 2.818 \times 10^{-15}}$ m, $\scriptstyle{t(\theta_{\rm in})}$ and $\scriptstyle{t(\theta_{\rm sc})}$ are the Fresnel transmission coefficients between water
and silicon, for the grazing angle of incidence $\scriptstyle{\theta_{\rm in}}$ and for the scattering
angle $\scriptstyle{\theta_{\rm sc}}$.  The coefficient $\scriptstyle{t(\theta_{\rm in})}$ represents a good
approximation to the actual field at the interface while $\scriptstyle{t(\theta_{\rm sc})}$ describes
how the scattered field propagates to the detector. 
Eq. (\ref{equagene}) has to be be multiplied by
the incident flux and (numerically) integrated over the detector solid angle to get the scattered intensity.
The three terms in between the square brackets describe the surface roughness and bilayers 1 and 2 static roughness and
thermal fluctuations respectively.
$\scriptstyle{\delta \tilde{\rho}_i(q_z)}$ (i=1,2) is the Fourier transform of the i-th bilayer (located at 
$\scriptstyle{z_i({\bf r_\parallel})}$) electron density profiles (form factors),
which are described using the so-called 1G-hybrid model \cite{Wiener89,malaquin10}.  
Expanding the square modulus in Eq. (\ref{equagene}), we get self- and cross- height-height
correlation functions of the substrate and bilayers, where the cross-correlations
are sensitive to the interaction potentials.\\ 
We describe the substrate correlation function using a self-affine correlation function \cite{sinha88}. 
Static and thermal correlation functions used in Eq. (\ref{equagene}) are derived in detail in Ref. \cite{malaquin10} using the free energy: 
\begin{eqnarray}
\tilde{{\cal F}}_q &=& \frac 12 \sum_{i=1}^2 \left[\left(\tilde{a}_i(q_\parallel) + U^{\prime\prime}_{12} \right) |\tilde{z}_{i}(q_\parallel)|^2 + U^{\prime\prime}_{is} \tilde{z}_{i}(q_\parallel) \tilde{z}_s(-q_{\parallel})\right] - U^{\prime\prime}_{12} \tilde{z}_{1}(q_\parallel) \tilde{z}_{2}(-q_\parallel),
\label{freetherm}
\end{eqnarray}
with $\scriptstyle{\tilde a(q_\parallel) = {{U}^{\prime\prime}_{is}+\gamma_i q_\parallel^2 +\kappa_i q_\parallel^4}}$,
where $\scriptstyle{\gamma_i}$ and $\scriptstyle{\kappa_i}$ are respectively the tension and the bending modulus of the i-th
bilayer.  $\scriptstyle{U^{\prime\prime}_{is}}$ and $\scriptstyle{U^{\prime\prime}_{12}}$ are second derivatives of the effective interaction
potential between the substrate and a bilayer and between bilayers. 
The linear response theory of Swain and Andelman \cite{andelman2001} was extended to double bilayers
in order to describe the static coupling of the bilayers to the substrate and
the thermal correlation functions were derived by diagonalizing $\scriptstyle{\tilde{{\cal F}}_q}$, applying the equipartition of energy 
and Fourier transforming.\\

\begin{acknowledgments}
We wish to thank J.-S. Micha for assistance during the experiments at ESRF (BM32), the ILL for support laboratories for sample preparation, P. K\'ekicheff and C. Marques for stimulating discussions.
\end{acknowledgments}


\newcommand{\SortNoop}[1]{}

\newpage

\begin{center}
{\bf \Large SUPPORTING INFORMATION}
\end{center}

\setcounter{section}{0}
\section{Fit parameters}

\begin{figure}[h!]
\includegraphics[width=0.5\textwidth]{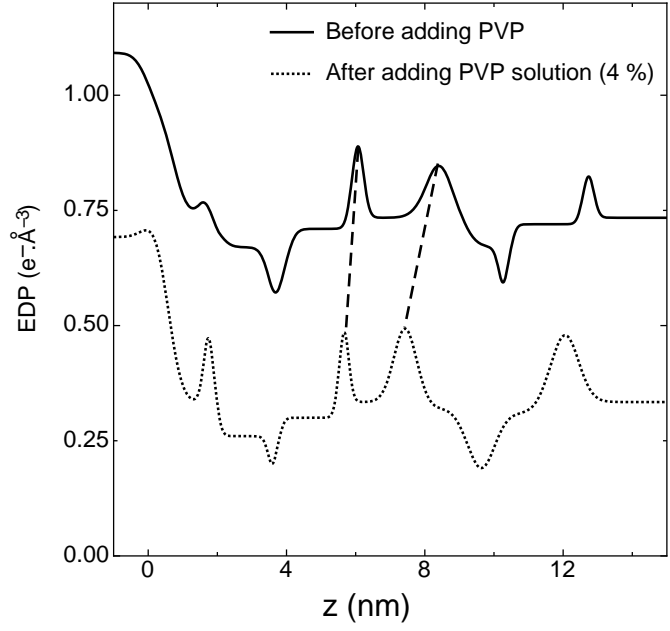}
\caption{Example of Electron Density Profiles showing the effect of osmotic pressure. The upper curve has been shifted of 0.4 e$^-$/\AA$^{3}$ for clarity.}
\label{EDP_quadri}
\end{figure}

As described in the manuscript, we perform combined fits of experimental specular and off-specular data. Best fit main parameters for elastic constants ($\gamma_2$, $\kappa_2$), interaction potential second derivative ($U^{\prime\prime}_{M_1,M_2}$), static ($\sigma_{st,2}$) and thermic ($\sigma_{th,2}$) roughnesses of the floating bilayer are given in Tables \ref{table_ots_kappa_gamma_u} and \ref{table_dspc_kappa_gamma_u}. A typical Electron Density Profile (EDP) for a double supported bilayer is also reported as Fig. \ref{EDP_quadri}.

\subsection{Influence of temperature and osmotic pressure}

\begin{table}[h]
\begin{center}
\begin{tabular}{|c||c c c c c||c c|}
	\hline
		 & \multicolumn{5}{c||}{Influence of temperature} & \multicolumn{2}{c|}{Osmotic pressure} \\
	\cline{2-8}
	& 42.9$^\circ$C & 49.7$^\circ$C & 51.5$^\circ$C & 52.0$^\circ$C & 52.9$^\circ$C & PVP 4\% & PVP 30\% \\
	\hline
	\hline
	$D_{2,H_2O}$ [\AA] & 25.0$\pm$0.2 & 26.5$\pm$0.2 & 26.9$\pm$0.2 & 28.1$\pm$0.2 & 29.7$\pm$0.2 & $18.0 \pm 1$ & $14.0 \pm 1$  \\
	\hline
	\hline
	$\kappa_2$ [k$_B$T] & 280$\pm$50 & 300$\pm$50 & 300$\pm$50 & 60$\pm$20 & 50$\pm$20 & 200$\pm$50 & 500$\pm$50  \\
	$\gamma_2$ [mN/m] & 0.3$\pm$0.2 & 0.3$\pm$0.2 & 0.4$\pm$0.2 & 0.4$\pm$0.2 & 0.5$\pm$0.2 & 1.11$\pm$ 0.3 & 0.3$\pm$0.1 \\
	$\sigma_{2,st}$ [\AA] & 2.0$\pm$0.5 & 2.5$\pm$0.5 & 3.5$\pm$0.5 & 3.0$\pm$0.5 & 3.0$\pm$0.5 & 3.5$\pm$0.5 & 2.5$\pm$0.5  \\
	$\sigma_{2,th}$ [\AA] & 6.3$\pm$0.7 & 8.6$\pm$0.9 & 10.7$\pm$0.8 & 12.1$\pm$2 & 13.2$\pm$3 & 3.8$\pm$1 & 2.8$\pm$1  \\
	\hline
	\hline
	$U^{\prime\prime}_{M_1,M_2}$ [10$^x$ J/m$^4$] & 12.05$\pm$0.3 & 11.8$\pm$0.3 & 11.6$\pm$0.3 & 11.7$\pm$0.3 & 11.5$\pm$0.3 & 13.85$\pm$0.1 & 14.5$\pm$0.1  \\
	\hline
\end{tabular}
\end{center}
\caption{Temperature evolution and osmotic pressure effect.}
\label{table_ots_kappa_gamma_u}
\end{table}

\subsection{Influence of salt (NaCl)}

\begin{table}[h]
\begin{center}
\begin{tabular}{|c||c | c  | c|| c| c|}
	\hline	
	 & \multicolumn{3}{c||}{Fluid phase} & \multicolumn{2}{c|}{Gel phase} 	\\
	\cline{2-6}
	& Before salt & $l_D$=0.5 nm & $l_D$=0.4 nm & Before salt & $l_D$=0.4 nm \\
	\hline
	\hline
	$D_{2,H_2O}$ [\AA] & 28.0 $  \pm 1$ & $ 24.6 \pm 1$ &$ 20.3 \pm 1$ & 24.3 $ \pm 1$ & $ 22.2 \pm 1$ \\
	\hline
	\hline 
	$\kappa_2$ [k$_B$T] & 80 $\pm$50 & 473 $\pm$50  & 232 $\pm$50 & 75 $\pm$10 & 175 $\pm$ 10 \\
	$\gamma_2$ [mN/m] & 0.80 $\pm$ 0.1 & 0.15$\pm$0.1 &  0.79 $\pm$ 0.1 & 0.69  $\pm$ 0.1 & 0.51$\pm$0.1 \\
	$\sigma_{2,st}$ [\AA] & 0.5$\pm$0.5 & 1.0 $\pm$0.5  &  2.5 $\pm$0.5  &  1.0 $\pm$0.5 & 1.0  $\pm$ 0.5 \\
	$\sigma_{2,th}$ [\AA] & 7.5 $\pm$1 & 4.9 $\pm$1 & 4.1 $\pm$ 1 & 8.0 $\pm$1 & 5.8 $\pm$1 \\
	\hline
	\hline
	$U^{\prime\prime}_{M_1,M_2}$ [10$^x$ J/m$^4$] & 12.1 $\pm$0.1 & 12.9 $\pm$0.1 & 13.3 $\pm$ 0.1 & 12.2  $\pm$0.3 & 13.0 $\pm$0.1 \\
	\hline
\end{tabular}
\end{center}
\caption{Effect of salt screening in fluid and gel phase.}
\label{table_dspc_kappa_gamma_u}
\end{table}

\clearpage
\subsection{Example of Electron Density Profile}

\section{Interactions : Lifshitz approach}

The Lifshitz theory describes the electromagnetic interactions between two continuous media in term of fluctuations of the electromagnetic field, summing over all frequencies. This approach is more general than the {\it simple} sum of pairwise interactions between atoms. For a review see \cite{SMahanty1976}.

\begin{figure}
\includegraphics[width=0.3\textwidth]{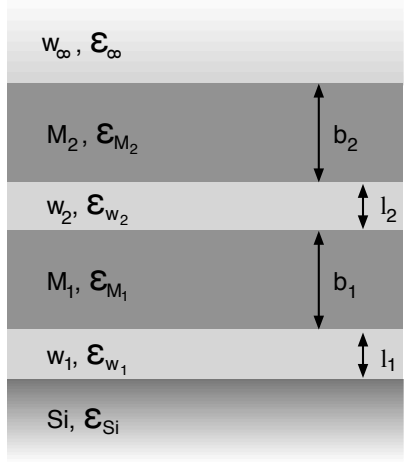}
\caption{Schematic representation of a supported double bilayer for the van der Waals potential calculation.}
\label{vdw_quadri}
\end{figure}

We assume a realistic model of supported double bilayers (see Fig. \ref{vdw_quadri}), with two semi-infinite media (water and silicon), two layers representing the two lipid bilayers (thicknesses $b_1$ and $b_2$) and two water layers (thicknesses $l_1$ and $l_2$). Lipid heads are highly hydrated and can be included in water layer. So in our case $l_2=d_w+2d_{head}$ where $d_{head}$ is the lipid head thickness. Using these model, Lifshitz theory leads to \cite{SMahanty1976}:

\begin{eqnarray}
	\nonumber
	U_{M_1,M_2}\left(l_2,T\right) = \frac{k_B T}{8\pi l_2^2} \sum_{n=0}^{\hspace{0.2cm}\infty \hspace{0.2cm} \prime} \int_{r_n}^{\infty} x \ln \left[\left(1-\overline{\Delta}_{Si,w_1}\left(l_1,b_1\right) \overline{\Delta}_{w_2,w_{\infty}}\left(b_2\right) e^{-x} \right) \right. \\
	\left. \times \left(1-\Delta_{Si,w_1}\left(l_1,b_1\right) \Delta_{w_2,w_{\infty}}\left(b_2\right) e^{-x} \right)\right] dx,
\end{eqnarray}

where $\overline{\Delta}$ and $\Delta$ depend on the different geometric parameters ($b_i, l_i$) and on dielectric constants $\epsilon_i$ of each medium. The full expression can be found in reference \cite{SMahanty1976}. The key point in such a calculation is to accurately describe the dielectric permittivity frequency dependence. Following \cite{SShubin1993}, $\epsilon_i\left(\omega\right)$ is given by ~:

\begin{eqnarray}
 	\epsilon_i(\omega) = 1 + \sum_r \frac{C_r}{1+i \omega/\omega_r} + \sum_{p} \frac{C_p}{1 - \omega^2/\omega_p^2 + i \gamma_p \omega/\omega_p^2}
\end{eqnarray}

The $\omega_{r,p}$ are the absorption frequencies, $C_{r,p}$ the absorption strength and the $\gamma_{r,p}$ are the damping factors. The first sum corresponds to microwave relaxation and the second one to experimental absorption peaks, usually corresponding to Infrared and Ultraviolet frequencies (see Table \ref{tab_hamaker}). As shown by Ninham and Parsegian \cite{Sninham(biophysj1970)}, in biologically relevant cases like lipid-water systems, it is of high importance to use the total dielectric data from component substances to analyze the fluctuations. In particular, infrared and microwave frequencies are very important as confirmed convincingly by Surface Force Apparatus experiments \cite{SShubin1993}.

 \begin{table}[h]
 \begin{center}
 \begin{tabular}{| c || c || c | c || c | c | c || c | c | c |}
 	\hline
	 \multicolumn{2}{|c||}{} & \multicolumn{2}{c||}{Microwave} & \multicolumn{3}{c||}{Infrared} & \multicolumn{3}{c|}{Ultraviolet} 	\\
	\hline
 	& $\epsilon_r$ &  $\omega_{MW}$ & $C_{MW}$ &  $\omega_{IR}$ & $C_{IR}$ & $\gamma_{IR}$ & $\omega_{UV}$ & C$_{UV}$ & $\gamma_{UV}$ \\
 	& & [$10^{11}$ rad.s$^{-1}$] & & [$10^{14} $rad.s$^{-1}$] & & [$10^{13}$ rad.s$^{-1}$] & [$10^{16}$ rad.s$^{-1}$] & & [$10^{15}$ rad.s$^{-1}$] \\
 	\hline
 	Silicon & 11.6 & & & 0.345 & 0.043 &  0 & 0.503 & 10.448 & 0 \\
	& & & & 0.535 & 0.050 & 0 & & &  \\
 	& & & & 0.884 & 0.059 & 0 & & &  \\
 	\hline
 	\hline
 	Water & 80.1 & 1.083 & 75.3 & 0.314 & 1.4635 & 2.29 & 1.2593 & 0.0392 & 0.774  \\
	& & & & 1.047 & 0.7368 & 5.78 & 1.5172 & 0.0570 & 1.335  \\ 
	& & & & 1.396 & 0.1526 & 4.22 & 1.7296 & 0.0923 & 2.336 \\
	& & & & 3.065 & 0.0136 & 3.81 & 1.9724 & 0.1556 &  3.110 \\ 
	& & & & 6.450 & 0.0751 & 8.54 & 2.2606 & 0.1522 & 4.491 \\
	& & & & & & & 2.8068 & 0.2711 & 9.498 \\
 	\hline
 	\hline
 	Alkanes & 2.014 &  &  & 5.54 & 0.025 & 0 & 1.848 & 1.026 & 0  \\
	\hline
 \end{tabular}
 \end{center}
 \caption{Values of the constants used for the dielectric response $\epsilon\left(\omega\right)$ of Silicon \cite{SSenden(colloid1995)}, Water \cite{SShubin1993} and \cite{Shunter(Book2001)}.}
 \label{tab_hamaker}
 \end{table}
 
 \begin{figure}[h!]
\includegraphics[width=0.5\textwidth]{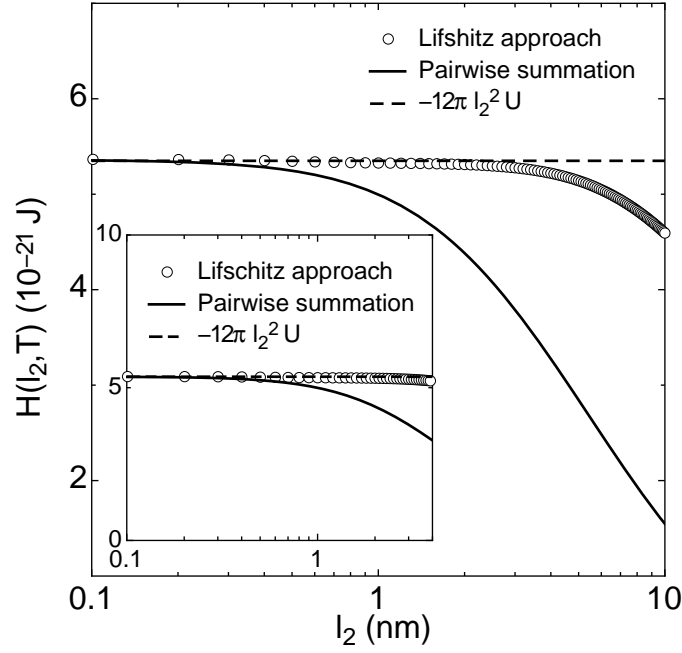}
\caption{Effective Hamaker constant vs $l_2$ for a Si/Water/Bilayer/Water/Bilayer/Water, pairwise summation (black solid line) and expression of the form $-H(l_2=0,T)/12\pi l_2^2$ (black dashed line). In inset, zoom on the range of interest for our experiments.}
\label{plot_hamaker_ots_bic}
\end{figure}

An effective Hamaker ``constant'' is then defined by	$H\left(l_2,T\right)= -12\pi l_2^2 U_{M_1,M_2}\left(l_2,T\right)$. For interlayer water thicknesses $d_w=l_2+2d_{head}$ larger than a nanometer, pairwise summation and Lifshitz theory give different values for the Hamaker constant (see Fig. \ref{plot_hamaker_ots_bic}). For our experimental values ($d_w < 3$ nm), the van der Waals interaction given by Lifshitz theory is well described by : $-H(l_2=0,T)/12\pi\left(d_w+2d_{head}\right)^2$, with $H(l_2=0,T) \sim 5.3\times10^{-21}$ J.

\section{Self-Consistent theory (SC) and Effective potential in asymetrical potential}
 \label{subsec_cas_membrane_supportee}
 
The renormalization of the microscopic interaction potential by the thermal fluctuations is a complex problem of modern statistical physic. Helfrich first realized that, in addition to the ``direct'' electrostatic, van der Waals and hydration
forces \cite{Sliporevue}, the long-range ``effective'' steric interaction generated by the thermal fluctuations 
of confined flexible membranes is an essential contribution to the total free energy of interaction \cite{Shelfrich73}.
The effective potential, average bilayer position, and fluctuation amplitude are thus coupled quantities which must be self-consistently determined. 
 
 \subsection{Effective potential theory}

Effective potential theories consist in adding an entropic (fluctuations) term $U_{\rm fl}\left(z\right)$ to the direct potential $U\left(z\right)$. The $d_w$ equilibrium value is thus given by the minimum of $U_{\rm tot}^{\prime}\left(d_w\right)=\left(U+U_{\rm fl}\right)^{\prime}\left(z=d_w\right)=0$. One can also compute the second derivative of the external potential at the equilibrium position  $U_{\rm tot}^{\prime\prime}\left(d_w\right)$. 

\begin{itemize}

\item Pure hard wall interaction (hard confinement) was first considered in \cite{Shelfrich73} leading to an entropic contribution~: 

\begin{equation}
U_{\rm fl}=U_{\rm Hel} = c_H /\kappa (k_BT/z)^2,
\end{equation}

with $c_H$ ranging from 0.08 to 0.2 \cite{SGompper1989,SMecke2003} in symmetrical case.

\item Podgornik and Parsegian have extended Helfrich's approach \cite{Shelfrich78} to take into account hydration repulsion and van der Waals attraction in the so-called  ``soft'' potential \cite{SPodgornik1992}, 

\begin{equation}
U_{\rm fl}=U_{\rm soft} = \pi k_B T /16 \sqrt{P_h/\kappa/z_h} \exp(-z/n z_{h}).
\end{equation}

\end{itemize}

\subsection{Self-consistent theory}
 
Self-consistent theory developed by  \cite{SMecke2003} considers the position fluctuations of a membrane close to a substrate in an external potential. It is an alternative approach to compute the average membrane position and its root mean square fluctuation amplitude in the range of moderate fluctuations much below the unbinding transition. For arbitrary potentials, partition function ${\cal Z}$, mean membrane-substrate distance $d_w$,  fluctuation amplitude r.m.s.  $\sigma_{th}$ and mean value of second derivative of the external potential can be expressed as self-consistent equations :

\begin{eqnarray}
	\nonumber
	{\cal Z} &=& \int dz \exp\left(- \frac{16 \kappa \sigma_{th}^2}{\left(k_B T\right)^2} U\left(z\right) - \frac{3\left(z-d_w\right)^2}{8\sigma_{th}^2}\right),\\
	\nonumber
	d_w &=& \langle z \rangle = \frac{1}{{\cal Z}} \int dz z \exp\left(- \frac{16 \kappa \sigma_{th}^2}{\left(k_B T\right)^2} U\left(z\right) - \frac{3\left(z-d_w\right)^2}{8\sigma_{th}^2}\right),\\
	\nonumber
	\sigma_{th}^2 &=& \langle \left(z-d_w\right)^2 \rangle = \frac{1}{{\cal Z}} \int dz \left(z-d_w\right)^2 \exp\left(- \frac{16 \kappa \sigma_{th}^2}{\left(k_B T\right)^2} U\left(z\right) - \frac{3\left(z-d_w\right)^2}{8\sigma_{th}^2}\right).\\
	\nonumber
	 \langle U^{\prime\prime} \rangle &=& \frac{1}{{\cal Z}} \int dz U^{\prime\prime}\left(z\right) \exp\left(- \frac{16 \kappa \sigma_{th}^2}{\left(k_B T\right)^2} U\left(z\right) - \frac{3\left(z-d_w\right)^2}{8\sigma_{th}^2}\right).
\label{eq_auto_coherente}
\end{eqnarray}

 \subsection{Comparison between both approaches}

\begin{figure}[h!]
\includegraphics[width=0.5\textwidth]{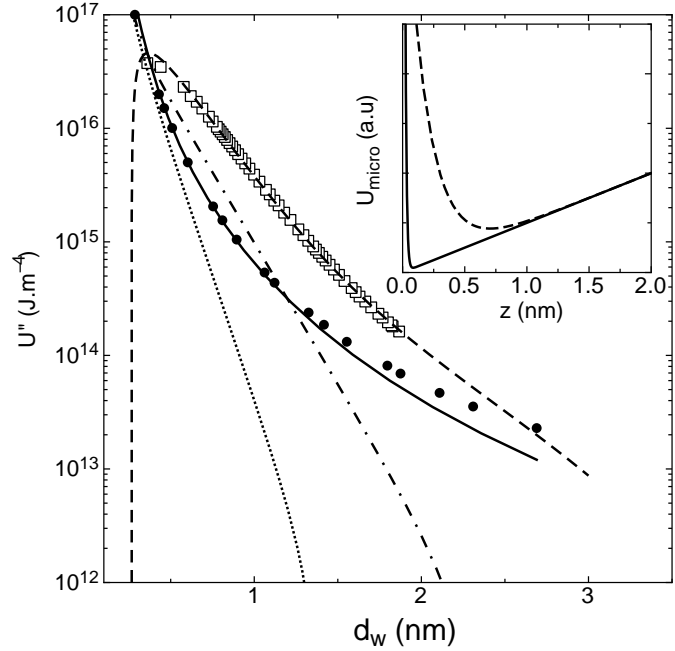}
\caption{Mean values of the external potential second derivative calculated using self-consistent approach (points) and using effective potential theory (lines) for soft-repulsion ($\square$ and dashed line ($P_h=0.5\cdot 10^7$ Pa, $z_h=0.18$ nm)) and for hard repulsion ($\bullet$ and solid line ($P_h=0.5\cdot 10^9$ Pa, $z_h=0.005$ nm)). The inset shows the associated external potential.}
\label{graph_self_consistant_helfrich}
\end{figure}

We have solved the self-consistent equations with an asymmetric external potential  $U\left(z\right)$ composed of hydration pressure term $U_{\rm hyd}(z) = P_h z_h \exp(-z/z_h)$ and an osmotic pressure $U_{\rm osmo}(z) = P z$ (see Fig. \ref{graph_self_consistant_helfrich} inset). By varying $P_h$ and $z_h$ from ($P_h=0.5\cdot 10^7$ Pa, $z_h=0.18$ nm) to ($P_h=0.5\cdot 10^9$ Pa, $z_h=0.005$ nm) we were able to mimic a transition from a soft hydration repulsion to hard wall steric repulsion.

By comparing the results obtained by the self-consistent and ``effective potential''  theories (Fig.\ref{graph_self_consistant_helfrich}) we observe that~: 

\begin{itemize}

\item In the case of a soft repulsion, self-consistent theory ($\square$) and effective potential theory (blue solid line) are in good agreement, in particular in the small fluctuation regime, and for various bending modulus ranging from 5 to 200$k_B T$ (not shown here). It is impossible to adjust a ``hard wall" model , whatever the bending modulus used. However, a perfect fit can be obtained with a ``soft potential" $U_{\rm hyd}+U_{\rm soft}$, where the only fit parameter is the screening coefficient $n$ in the confinement potential $U_{\rm soft}(z) \propto \exp(-z/nz_h)$. The value obtained for $n$ is $n=2.4$, in very good agreement with the literature \cite{Spetrache(pre1998)}. Nevertheless, the self-consistent approach is underestimating the mean distance $d_w$ and overestimating $\langle U^{\prime\prime} \rangle$.

\item In the case of hard wall repulsion ($\bullet$), it is only possible to fit the results with Helfrich effective potential: $U_{\rm Hel} = c_H /\kappa (k_BT/z)^2$, with $c_H \sim 0.5$. Whatever the parameters used, a ``soft potential" model fails to describe this hard wall system. The value of $c_H$ is twice as large as the values obtained in the symmetrical case.

\end{itemize}

These results show that self-consistent theory is able to describe the continuous transition from a soft hydration repulsion to an hard steric one.

\newpage
\clearpage

\end{document}